\documentclass[
 reprint,
 aps,
 prl,
 superscriptaddress,
 showpacs,
 amsmath,
 amssymb,
]{revtex4-2}

\usepackage{graphicx}
\usepackage{dcolumn}
\usepackage{bm}
\usepackage{mathbbol}

\begin{document}

\preprint{APS/123-QED}

\title{A Measurement of the Effective Mean-Free-Path of Solar Wind Protons}

\author{Jesse T. Coburn}
\author{Christopher H. K. Chen}%
\affiliation{Department of Physics and Astronomy, Queen Mary University of London, London E1 4NS, United Kingdom}

\author{Jonathan Squire}
\affiliation{Physics Department, University of Otago, Dunedin 9010, New Zealand}

\date{\today}

\begin{abstract}
Weakly collisional plasmas are subject to nonlinear relaxation processes, which can operate at rates much faster than the particle collision frequencies. This causes the plasma to respond like a magnetised fluid despite having long particle mean-free-paths. In this Letter the \emph{effective} collisional mechanisms are modelled in the plasma kinetic equation to produce density, pressure and magnetic field responses to compare with spacecraft measurements of the solar wind compressive fluctuations at 1 AU. This enables a measurement of the effective mean-free-path of the solar wind protons, found to be 4.35 $\times 10^5$ km, which is $\sim 10^3$ times shorter than the collisional mean-free-path. These measurements are shown to support the effective fluid behavior of the solar wind at scales above the proton gyroradius and demonstrate that effective collision processes alter the thermodynamics and transport of weakly collisional plasmas.
\end{abstract} 

\maketitle

\emph{Introduction}.---
Many natural plasmas (e.g., interstellar medium, galaxy clusters, black hole accretion disks, solar wind) are in a weakly collisional state, where the particle collision frequency $\nu_{\mathrm{coll}}$ is smaller than other characteristic frequencies (e.g., proton gyrofrequency $\Omega_{\mathrm{p}}$, inverse magnetic field correlation time $1/\tau_c$ etc.)  \citep{schekochihin2009astrophysical, marsch2006kinetic, schekochihin2006turbulence, quataert2003radiatively}. Thus, the characteristic timescales of the plasma motions $\omega$ can span from collisional (fluid) $\omega \ll \nu_{\mathrm{coll}}$ to collisionless $\omega \gg \nu_{\mathrm{coll}}$ \citep{schekochihin2009astrophysical, schekochihin2005plasma}. Knowledge of the transition scale $\omega \sim \nu_{\mathrm{coll}}$ is vital to understand the thermodynamics of astrophysical plasmas \citep{schekochihin2009astrophysical}.

The escaping solar corona, known as the solar wind, expands into interplanetary space as a super-Alfv\'{e}nic and turbulent plasma  \citep{parker1958dynamics, bruno2013solar, verscharen2019multi}. In situ measurements of particle distribution functions and electromagnetic fields enable fundamental plasma physics observations \citep{chen2016recent}. The Spitzer-H\"{a}rm proton-proton collision frequency $\nu_{\mathrm{p},\mathrm{p}}^{\mathrm{SH}}$ decreases with radial distance from the Sun, and by a few solar radii, is much smaller than other characteristic frequencies. In principle, the dynamics should be described by collisionless plasma equations. For reference, at 1 AU, typical frequencies are $\nu_{\mathrm{p},\mathrm{p}}^{\mathrm{SH}} \approx 4 \times 10^{-7} \, \mathrm{s}^{-1}$, $\Omega_{\mathrm{p}} \approx 10^{-1} \, \mathrm{s}^{-1}$, $1/\tau_c \approx  10^{-6} \, \mathrm{s}^{-1}$ \,\citep{book1987nrl, spitzer2006physics, verscharen2019multi, matthaeus2010eulerian}. 
 
Despite the weak collisionality of the solar wind, many aspects appear to be described by fluid equations: magnetohydrodynamic (MHD) turbulence theory predicts the shape of power spectra (e.g., magnetic field, proton density) \citep{coleman1968turbulence, matthaeus1982measurement, bruno2013solar, goldreich1997magnetohydrodynamic, tu1995mhd}, spatial transport \citep{matthaeus1999turbulence, zank1996evolution}, and the proton heating rate by the energy cascade \citep{macbride2008turbulent, stawarz2009turbulent, coburn2012turbulent}. While this is due, in part, to the dominance of Alfv\'{e}nic fluctuations \citep{schekochihin2009astrophysical, lithwick2001compressible, howes2008amodel, quataert1998heating}, compressive fluctuations, that should be severely damped in a collisionless plasma \citep{barnes1966collisionless}, display the MHD slow-mode polarization (magnetic and thermal pressure anticorrelated). They are routinely detected at a range of scales \citep{tu1995mhd, howes2012slow, klein2012using, kellogg2005rapid, yao2013small_FH, yao2013small_SM, yao2011multi} following a power law predicted from the MHD equations \citep{montgomery1987density, schekochihin2009astrophysical, lithwick2001compressible, marsch1990spectral}. The most clear evidence of fluid behavior can be seen in the strong correlation between the density and thermal pressure, indicating a polytropic equation of state \citep{marsch1983equation, 2020ApJ...901...26N, totten1995empirical, verscharen2017kinetic}.

While the Spitzer-H\"{a}rm collision frequency appears incompatible with the fluidlike behavior of the solar wind, weakly collisional plasmas are also subject to nonlinear processes that prevent extreme departure from equilibrium \citep{nishida1969thermal, griffel1969anisotropy, yoon2017kinetic, hamasaki1973relaxation, gary2000electromagnetic}. Solar wind observations present substantial evidence of temperature anisotropy instabilities constraining the particle distribution functions \citep{kasper2002wind, hellinger2006solar, chen2016multi, marsch2006kinetic, tu2002anisotropy, yoon2017kinetic} suggesting they experience pitch-angle scattering by plasma waves \citep{bale2009magnetic}. These processes can play a similar role to collisions i.e., they are \emph{effective} collision processes.

This Letter presents a measurement of the effective mean-free-path of the solar wind by comparing observations of compressive wave-mode polarizations to numerical solutions of varying effective collisionality. It is shown that the transition from fluid to collisionless dynamics in the solar wind occurs at scales several orders of magnitude below the classical Spitzer-H\"{a}rm mean-free-path, explaining the fluidlike behavior of the weakly collisional solar wind.
  
\emph{Theory and numerical solutions}.--- The kinetic MHD equations with the Bhatnagar-Gross-Krook (BGK) collision operator \citep{bhatnagar1954a, gross1956model} produce dispersion relations and plasma fluctuations (e.g., magnetic field and pressure) that span between the collisionless and collisional limits \citep{snyder1997landau, sharma2003transition, kulsrud1983mhd}. They describe a nonrelativistic, magnetized plasma of arbitrary collision frequency \citep{kulsrud1983mhd, sharma2003transition}. The specific equations, which we refer to as KMHD-BGK, model both the proton and electron responses with the kinetic equation.  The BGK operator is used here to model relaxation processes, not particle collisions, so we use the language of an \textit{effective} proton collision frequency $\nu_{\mathrm{eff}}$ or mean-free-path $\lambda_{\mathrm{mfp}}^{\text{eff}} = v_{\text{th}}^{\mathrm{p}} / \nu_{\mathrm{eff}}$, where the proton thermal speed is $v_{\text{th}}^{\mathrm{p}}$ (see Supplemental Material).

Assuming plasma motions are slow compared to the gyrofrequency $\Omega_p$, the second moment of the kinetic equation and the ideal induction equation leads to,
\begin{subequations}
\label{eq:cgl}
\begin{align}
n_{\mathrm{p}} B \frac{d}{dt} \bigg( \frac{p_{\perp}^{\mathrm{p}}}{n_{\mathrm{p}} B} \bigg) &= - \bm{\nabla} \cdot ( q^{\mathrm{p}}_{\perp} \hat{\bm{b}})
\nonumber \\
& - q^{\mathrm{p}}_{\perp} \bm{\nabla} \cdot \hat{\bm{b}} + \frac{\nu_{\text{eff}}}{3} (p_{\parallel}^{\mathrm{p}} - p_{\perp}^{\mathrm{p}}),
\\
\frac{n_{\mathrm{p}}^3}{2 B} \frac{d}{dt} \bigg( \frac{p_{\parallel}^{\mathrm{p}} B^2}{n_{\mathrm{p}}^3} \bigg) &= - \bm{\nabla} \cdot (q_{\parallel}^{\mathrm{p}} \hat{\bm{b}} ) 
\nonumber \\
&+ q_{\perp}^{\mathrm{p}} \bm{\nabla} \cdot \hat{\bm{b}} + \frac{2 \nu_{\text{eff}} }{3} ( p_{\perp}^{\mathrm{p}} - p_{\parallel}^{\mathrm{p}} ), 
\end{align}
\end{subequations}
where $d/dt$ is the convective derivative and the quantities are the proton density $n_{\mathrm{p}}$, magnetic-field strength $B$, (perpendicular) parallel $(p_{\perp}^{\mathrm{p}}) \; p_{\parallel}^{\mathrm{p}}$ proton pressure, field parallel flux of (perpendicular) parallel $ (q_{\perp}^{\mathrm{p}}) \; q_{\parallel}^{\mathrm{p}}$ proton heat, and the unit magnetic field vector $\hat{\bm{b}} = \bm{B} / B$  \citep{chew1956boltzmann, hunana2019introductory}. The Alfv\'{e}n speed is $v_A= B/\sqrt{4 \pi n_{\mathrm{p}} m_{\mathrm{p}}}$, the proton gyroradius is $\rho_{\mathrm{p}} = v_{\mathrm{th}}^{\mathrm{p}} / \Omega_{\mathrm{p}}$, and the ion-acoustic speed is $c_{\mathrm{s}} = \sqrt{(3 k_{\mathrm{B}} T_{\parallel}^{\mathrm{p}} + k_{\mathrm{B}} T_{\parallel}^{\mathrm{e}})/m_{\mathrm{p}}}$, where the parallel proton (electron) temperature is $T_{\parallel}^{\mathrm{p}} \; (T_{\parallel}^{\mathrm{e}})$.

Equations \eqref{eq:cgl} are often discussed when the right hand sides are zero and are then referred to as the double adiabatic equations or Chew-Goldberger-Low (CGL) invariants \citep{chew1956boltzmann}. The focus here is on how the CGL invariants are broken, for example, by the heat flux terms in the collisionless limit, and by the collisional terms ($\propto \nu_{\mathrm{eff}}$). Therefore, the relative conservation of the CGL invariants provides a sensitive test of the equation of state \citep{schekochihin2010magnetofluid}. 

We proceed by constructing measures that describe the correlation and amplitude ratios of the left hand sides of Eqs. \eqref{eq:cgl},
\begin{subequations}
\label{eq:corrs_amps}
\begin{eqnarray} 
C_{\parallel}  &= \frac{ \langle \delta p_{\parallel}^{\mathrm{p}}\, \delta (n^3_{\mathrm{p}} / B^2) \rangle }{ \langle |  \delta  p_{\parallel}^{\mathrm{p}}  |^2 \rangle^{1/2} \langle | \delta (n^3_{\mathrm{p}} / B^2) |^2 \rangle^{1/2}} \label{subeq:c_par},
\\
A_{\parallel}  &= \frac{ \langle | \delta (n^3_{\mathrm{p}}/ B^2) |^2 \rangle^{1/2} }{ \langle  n^3_{\mathrm{p}}/ B^2  \rangle }  \frac{ \langle  p_{\parallel}^{\mathrm{p}}  \rangle }{ \langle | \delta p_{\parallel}^{\mathrm{p}} |^2 \rangle^{1/2} } \label{subeq:a_par},
\\
C_{\perp} &= \frac{ \langle \delta p_{\perp}^{\mathrm{p}} \, \delta (n_{\mathrm{p}} B) \rangle }{ \langle | \delta p_{\perp}^{\mathrm{p}} |^2 \rangle^{1/2} \langle | \delta (n_{\mathrm{p}} B) |^2 \rangle^{1/2}} \label{subeq:c_perp},
\\
A_{\perp}  &= \frac{ \langle | \delta (n_{\mathrm{p}} B) |^2 \rangle^{1/2} }{ \langle  n_{\mathrm{p}} B  \rangle }  \frac{ \langle  p_{\perp}^{\mathrm{p}}  \rangle }{ \langle | \delta p_{\perp}^{\mathrm{p}} |^2 \rangle^{1/2} } \label{subeq:a_perp},
\end{eqnarray}
\end{subequations}
where $\delta \chi = \chi - \langle \chi \rangle$ is the fluctuation about the average $\langle \chi \rangle$. The method compares predictions for Eqs. \eqref{eq:corrs_amps} derived from the slow-mode eigenmodes of the linearized KMHD-BGK system (e.g., $\delta p_{\perp}^{\mathrm{p}}, \, \delta B$ etc.) to solar wind measurements. See Supplemental Material.

The model's free parameters are the propagation angle $\theta_{\hat{\bm{b}},\hat{\bm{k}}}$ and proton effective mean-free-path $\lambda_{\mathrm{mfp}}^{\mathrm{eff}}$, i.e., they can be measured. The wavenumber $k$ and the proton beta $\beta =   8 \pi k_B p^{\mathrm{p}} / B^2$ where $p^{\mathrm{p}} = 2p^{\mathrm{p}}_{\perp}/3 + p^{\mathrm{p}}_{\parallel}/3$ are set to measured values. The species temperature ratio is set to a typical value for the solar wind $T_{\mathrm{p}} / T_{\mathrm{e}} = 1$ and the effective mean-free-path species ratio is set to $\lambda_{\mathrm{mfp}}^{\mathrm{eff}} / \lambda_{\mathrm{mfp,electrons}}^{\mathrm{eff}} = 5$; see Supplemental Material at for details of electron physics.

\begin{figure}
\includegraphics[width=86mm]{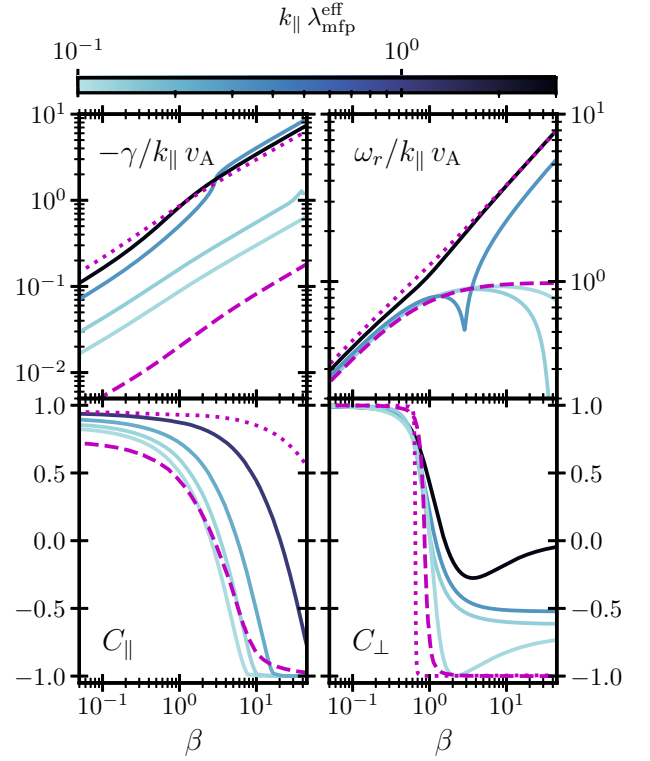}
\caption{\label{fig:disp} Numerical solutions of the KMHD-BGK equations for a range of $k_{\parallel} \lambda_{\mathrm{mfp}}^{\mathrm{eff}}$ (see color bar). The remaining parameters are defined in the text. Vertical axis labels are annotated on the panels, where $\gamma \; (\omega_r)$ are the imaginary (real) part of the complex frequency. The bottom panels are Eqs. \eqref{subeq:c_par}, \eqref{subeq:c_perp}. The dotted (dashed) magenta lines are the long (short) limit of $\lambda_{\mathrm{mfp}}^{\mathrm{eff}}$ corresponding to the collisionless (collisional) slow-mode / ion-acoustic branch for $\theta_{\hat{\bm{b}},\hat{\bm{k}}} = 88^{\circ}$\citep{verscharen2017kinetic, howes2006astrophysical}.}
\end{figure} 

The top panels of Fig. \ref{fig:disp} demonstrate the ability of the KMHD-BGK equations to resolve the dynamics of the compressive slow-mode from collisional (lighter blue) to collisionless (black) \citep{sharma2003transition}. Numerical predictions for Eqs. \eqref{subeq:c_par}, \eqref{subeq:c_perp} (bottom panels of Fig. \ref{fig:disp}) show distinct differences at $\beta > 1$ for $k_{\parallel} \lambda_{\mathrm{mfp}}^{\mathrm{eff}}$, which can be compared to observations. The MHD/collisionless limits are illustrated in magenta,  for $C_{\perp}$ (bottom right panel) these two limits produce similar trends, therefore it is necessary to make comparisons at multiple $k_{\parallel}$, to measure $\lambda_{\mathrm{mfp}}^{\mathrm{eff}}$.

\begin{figure}
\includegraphics[width=86mm]{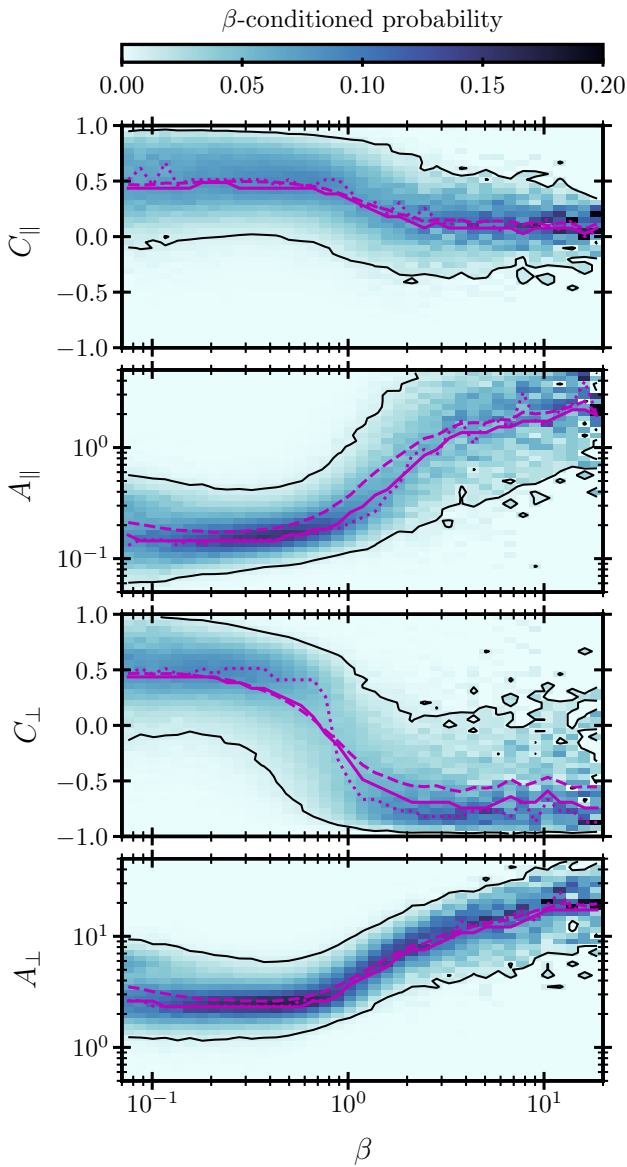}
\caption{\label{fig:correlations_amps} $\beta$-conditioned probability functions of the quantities in Eqs. \eqref{eq:corrs_amps} for the wavenumber bin $k_{\mathrm{SW}} = 0.288 \times 10^{-5} \, \mathrm{km}^{-1}$. The thin black line is a contour of probability equal to 0.01. The magenta lines are mean (dashed), median (solid), and mode (dotted) conditioned on the $\beta$-bins. }
\end{figure} 

\emph{Measurements}.---
The dataset consists of Wind spacecraft measurements of the pristine solar wind during years 2005-2010. The electrostatic analyzer 3DP records onboard moments of the proton density, velocity and pressure tensor, and the magnetometer MFI records the magnetic field, at a nominal $\sim 3s$ cadence \citep{lin1995three, lepping1995wind}.

The dataset is restricted to time intervals satisfying three criteria: (i) 95\% of the data is available (the remaining is then linearly interpolated); (ii) the median density must be greater than 1 particle per cm$^{-3}$; and (iii) the average norm of the non-gyrotropic tensor ($\mathsf{\Pi}^{\mathrm{p}}  = \mathsf{p}^{\mathrm{p}} -  \hat{\bm{b}} \hat{\bm{b}} \, p^{\mathrm{p}}_{\parallel} - ( \mathsf{1} - \hat{\bm{b}} \hat{\bm{b}} ) p_{\perp}^{\mathrm{p}}$), must be less than 30\% of the average norm of the pressure tensor $\mathsf{p}^{\mathrm{p}}$.

To probe a set of wavenumbers we measure the four quantities in Eqs. \eqref{eq:corrs_amps}, the average radial solar wind velocity $ \langle V_{\mathrm{SW}} \rangle $, and average proton beta for a set of time intervals $\tau = [30\text{s}, \, 1\text{min.},\,  2\text{mins.}, \, ... , \, 128\text{mins.}]$. The time scales are converted to wavenumber $k_{\mathrm{SW}} = 1/ \tau \langle V_{\mathrm{SW}} \rangle $ via Taylor's frozen-in-flow (TFF) assumption \citep{taylor1938spectrum}. Three bins of equal probability density are obtained where the median of each bin is $k_{\mathrm{SW}} = [0.288, 1.41, 6.34] \times 10^{-5} \, \mathrm{km}^{-1}$, which lie within the inertial range of the magnetic field power spectrum at 1 AU \citep{kiyani2015dissipation}. The wavenumber bins contain $[2.98, 16.6, 70.0] \times 10^5$ samples.

For bin $k_{\mathrm{SW}} = 0.288 \times 10^{-5} \, \mathrm{km}^{-1}$ the $\beta$-conditioned probability functions of Eqs. \eqref{eq:corrs_amps}, all mapped to a common color bar, are displayed in Fig. \ref{fig:correlations_amps}. The $\beta$-trend lines in magenta (see caption) capture statistically significant differences between $\beta \lessgtr 1$. From the correlations $C_{\parallel}, C_{\perp}$ it is clear that the CGL invariants are rarely conserved $(C_{\parallel}, C_{\perp} = 1)$, but display similar trends to the theoretical expectations (Fig. \ref{fig:disp}). The amplitude ratios $A_{\parallel}, A_{\perp}$ demonstrate a relative decrease in fluctuation amplitude of the pressure components at $\beta  > 1$.

\emph{Comparison of measurements and numerical solutions}.--- \label{sec:comp}
To compare theoretical predictions of Eqs. \ref{eq:corrs_amps} to observations, a degeneracy in parametrization must be dealt with: the numerical solutions primarily depend on $k_{\parallel}  \lambda_{\mathrm{mfp}}^{\mathrm{eff}} = k \, \mathrm{cos}\big(\theta_{\hat{\bm{b}},\hat{\bm{k}}}\big) \lambda_{\mathrm{mfp}}^{\mathrm{eff}}$ \citep{sharma2003transition}; such that $\lambda_{\mathrm{mfp}}^{\mathrm{eff}},\,\theta_{\hat{\bm{b}},\hat{\bm{k}}}$ are degenerate. To address this we introduce a scale dependent anisotropy model $(k_{\parallel} \sim k_{\perp}^{\alpha})$, which relates $k$ and $\theta_{\hat{\bm{b}},\hat{\bm{k}}}$,
\begin{align} \label{eq:k_anisotropy}
k =  \frac{k_{\mathrm{iso}}}{\sqrt{2}} \, \bigg[ \text{sin} \big( \theta_{\hat{\bm{b}},\hat{\bm{k}}} \big) \bigg]^{\alpha/(1-\alpha)}  \,\bigg[ \text{cos} \big(\theta_{\hat{\bm{b}},\hat{\bm{k}}} \big) \bigg]^{1/(\alpha - 1)},
\end{align}
where $k_{\mathrm{iso}}$ is the isotropic wavenumber ($k_{\perp} = k_{\parallel}$) and $\alpha$ is the anisotropy exponent, generalised from turbulence models \citep{goldreich1995toward}. Comparing solutions parameterized by $\lambda_{\mathrm{mfp}}^{\mathrm{eff}}, \, \alpha, \, k_{\mathrm{iso}}$ across multiple wavenumbers $k = k_{\mathrm{SW}}$ clears the degeneracy.

Finally, the predictions of Eqs. \eqref{eq:corrs_amps} from the numerical solutions are normalized to the measured $\beta$-conditioned mean value (dashed magenta lines in Fig. \ref{fig:correlations_amps}) of $C_{\parallel}, C_{\perp}, A_{\parallel}, A_{\perp}$ at $\beta \simeq 10^{-1}$. This is to account for the fact that linear polarizations are only approximately observed in strong turbulence \citep{chen2016recent}.

Ranges of  $\alpha, \, k_{\mathrm{iso}}, \, \lambda_{\mathrm{mfp}}^{\mathrm{eff}} = [0.05,1.0], \, [5 \times 10^{-9}, 5 \times 10^{-7}] \, \mathrm{km}^{-1}, \, [3.5 \times 10^4, 2.1 \times 10^6] \, \mathrm{km}$ are chosen for computing numerical solutions. The ranges of $\alpha, k_{\mathrm{iso}}$ are consistent with previous observations \citep{chen2012three, chen2016recent}. The range of $\lambda_{\mathrm{mfp}}^{\mathrm{eff}}$ returns numerical solutions of Eqs. \eqref{eq:corrs_amps} that compare qualitatively well with the observations (seen in Fig. \ref{fig:correlations_amps}). The Spitzer-H\"{a}rm mean-free-path returns the (collisionless) ion-acoustic dispersion relation which is inconsistent with the measurements.

To make a quantitative comparison, we compute the ``goodness of fit",
\begin{eqnarray} \label{eq:r_value}
R = \sqrt{N^{-1} \sum_i^N \, (\bar{y}_i - \hat{y}_i)^2},
\end{eqnarray}
where $\hat{y}_i$ ($\bar{y}_i$) is the local numerical solution (local measured mean), summed over $i$, denoting the $i$th $\beta$-bin. $R(k_{\mathrm{SW}}; \alpha, k_{\mathrm{iso}}, \lambda_{\mathrm{mfp}}^{\mathrm{eff}})$ is calculated for each wavenumber $k_{\mathrm{SW}}$, where the mean $\bar{y}_i$ is respective to the wavenumber bin. The $R$-values are inverted for unnormalized weights ($w = R^{-1}$), divided by the maximum weight, then summed over wavenumber $\mathcal{W} = \sum_k w(k; \alpha, k_{\mathrm{iso}}, \lambda_{\mathrm{mfp}}^{\mathrm{eff}})$ to break the aforementioned degeneracy. From this volume, weighted geometric means $\mu_x$, covariances $\sigma^2_{x,y}$, and two sigma confidence intervals $\mathrm{CI}_{x}$ are calculated (see Supplemental Material) \citep{kendall1946advanced, norris1940standard}. This is the method we employ to measure the quantities $\lambda_{\mathrm{mfp}}^{\mathrm{eff}}, \alpha, k_{\mathrm{iso}}$, providing the main results of the Letter.

\begin{figure}
\includegraphics[width=86mm]{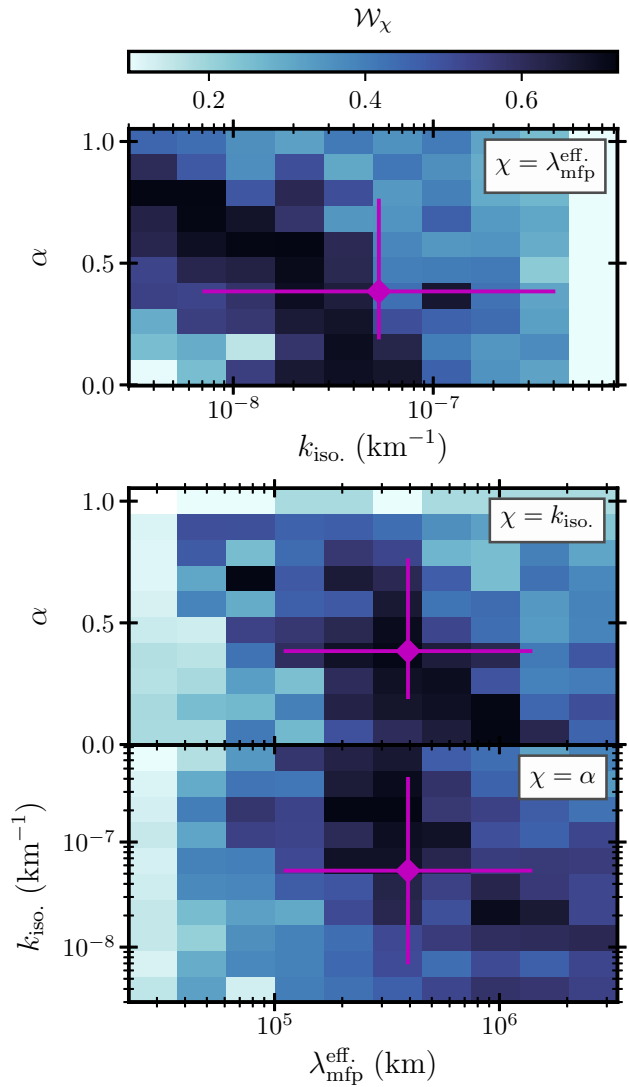}
\caption{\label{fig:int_plot} The three panels display the integrated ``goodness of fit" (Eq. \eqref{eq:int_weight}) for each parameter $\chi$, for $C_{\perp}$. The magenta crosses indicate the weighted geometric means and two sigma confidence intervals.}
\end{figure}

To visualize the weighted parameter space for $C_{\perp}$, Fig. \ref{fig:int_plot} illustrates the weight volume $\mathcal{W}(\alpha, k_{\mathrm{iso}}, \lambda_{\mathrm{mfp}}^{\mathrm{eff}})$, numerically integrated over each parameter axis $\chi$,
\begin{align} \label{eq:int_weight}
\mathcal{W}_{\chi} = \int_{\chi_0}^{\chi_n} \, d \chi \, \frac{\mathcal{W}(\alpha, k_{\mathrm{iso}}, \lambda_{\mathrm{mfp}}^{\mathrm{eff}})}{\chi_n - \chi_0},
\end{align}
where $\chi_n, \chi_0$ are limits of the range. The weighted means in Fig. \ref{fig:int_plot} lie in the maximum regions of $\mathcal{W}_{\chi}$, within the confidence intervals, indicating the weighted geometric statistics are a good representation of the observations. 

\begin{figure}
\includegraphics[width=86mm]{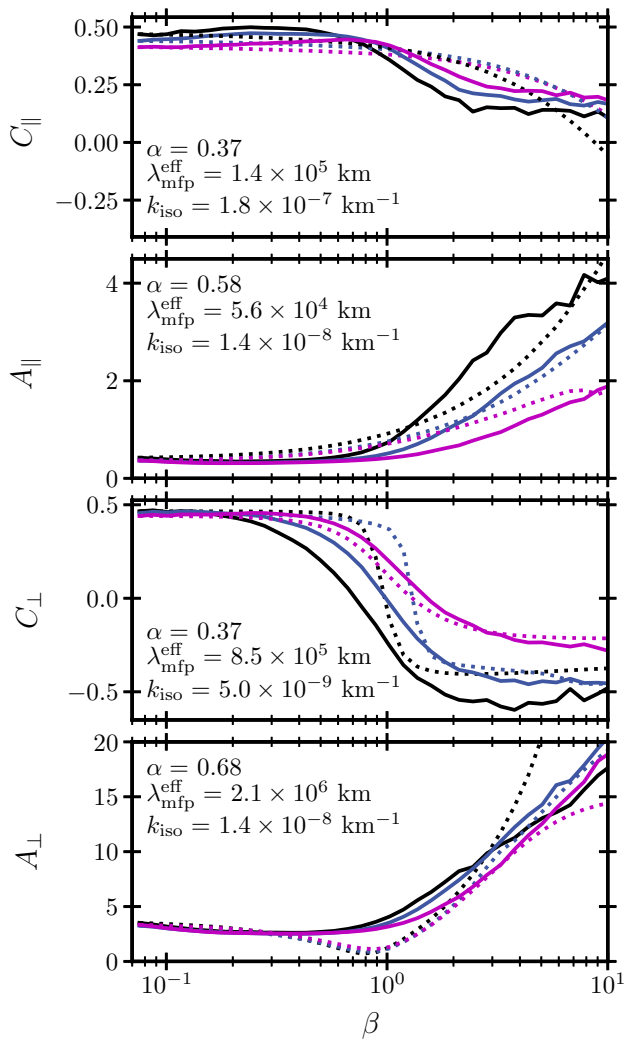}
\caption{\label{fig:multi_scale_best_fit} The four panels show the $\beta$-conditioned mean of the four quantities in Eqs. \eqref{eq:corrs_amps} for the three median wavenumber bins $k_{\mathrm{SW}} = [0.288, 1.41, 6.34] \times 10^{-5} \, \mathrm{km}^{-1}$ as solid (black, blue, magenta) lines respectively. Statistical uncertainties on the mean trends can be seen in Fig. \ref{fig:correlations_amps}. The dashed lines are the numerical solutions corresponding to the maximum $\mathcal{W}$; the parameters of the maxima are reported in the panels.}
\end{figure}

To check the scale dependence, Fig. \ref{fig:multi_scale_best_fit} displays the observed $\beta$-conditioned means of Eqs. \eqref{eq:corrs_amps} and the numerical solutions corresponding to the maximum $\mathcal{W}$. The numerical solutions and observations trend similarly with wavenumber indicating the scale dependence of the effective collisionality has been well modeled. The parameters of the maxima (recorded in the panels of Fig. \ref{fig:multi_scale_best_fit}) do not correspond exactly to the weighted geometric means of $\mathcal{W}$ (seen in Fig. \ref{fig:int_plot}) reflecting the statistical nature of the measured quantities.

The method of calculating statistics for $\lambda_{\mathrm{mfp}}^{\mathrm{eff}}, \, \alpha, \, k_{\mathrm{iso}}$ displayed in Fig. \ref{fig:int_plot} for $C_{\perp}$ produces similar statistics for $C_{\parallel}, A_{\parallel}, A_{\perp}$ (see Supplemental Material). Therefore, in Table \ref{tab:table1} combined statistics are reported. The measured effective mean-free-path and mean proton thermal speed (measured with this data set) gives an effective collision frequency of $\nu_{\mathrm{eff}} = v_{\mathrm{th}}^{\mathrm{p}} / \lambda_{\mathrm{mfp}}^{\mathrm{eff}}$ = 1.11 $\times 10^{-4}$ s$^{-1}$.
\begin{table}[b]
\caption{\label{tab:table1}%
Combined weighted geometric mean $\mu_x$, standard deviation $\sigma_{x,x}$, and the two sigma confidence interval $\mathrm{CI}_{x}$.
}
\begin{ruledtabular}
\begin{tabular}{lll}
\textrm{Statistic}&
\textrm{Value(s)} &
\textrm{Unit} \\
\colrule
$\mu_{\alpha} $&  0.425  & -
\\
$\mu_{k_{\mathrm{iso}}} $ & 5.36 $\times 10^{-8}$  & km$^{-1}$ 
\\
$\mu_{ \lambda_{\mathrm{mfp}}^{\mathrm{eff}}}$& 4.35 $\times 10^5$   & km 
\\
$\mathrm{CI}_{\alpha} $& [0.210, 0.858] & -
\\
$\mathrm{CI}_{k_{\mathrm{iso}}} $ & [0.0637, 4.51] $\times 10^{-7}$  & km$^{-1}$ 
\\
$\mathrm{CI}_{ \lambda_{\mathrm{mfp}}^{\mathrm{eff}}}$ & [0.102, 18.6] $\times 10^5$   & km 
\\
$\sigma^2_{\alpha,k_{\mathrm{iso}}} / (\sigma_{\alpha, \alpha} \, \sigma_{k_{\mathrm{iso}}, k_{\mathrm{iso}}})$&  0.221 & -
\\
$\sigma^2_{\alpha,  \lambda_{\mathrm{mfp}}^{\mathrm{eff}}} /   (\sigma_{\alpha, \alpha} \, \sigma_{ \lambda_{\mathrm{mfp}}^{\mathrm{eff}},  \lambda_{\mathrm{mfp}}^{\mathrm{eff}}}) $& 0.336 & - 
\\
$\sigma^2_{k_{\mathrm{iso}},  \lambda_{\mathrm{mfp}}^{\mathrm{ef f}}} / (\sigma_{k_{\mathrm{iso}}, k_{\mathrm{iso}}} \sigma_{ \lambda_{\mathrm{mfp}}^{\mathrm{eff}},  \lambda_{\mathrm{mfp}}^{\mathrm{eff}}})$ & 0.172 &-
\end{tabular}
\end{ruledtabular}
\end{table}

The transition frequency, where $\nu_{\mathrm{eff}} \simeq \omega$ can be estimated with $\nu_{\mathrm{eff}} = v_{\mathrm{th}}^{\mathrm{p}} / \lambda_{\mathrm{mfp}}^{\mathrm{eff}} $ and the ion-acoustic dispersion relation $\omega_{\mathrm{IA}} = k_{\parallel} \, c_{\mathrm{s}}$, giving the parallel transition wavenumber $k^{\mathrm{trans}}_{\parallel} = v_{\mathrm{th}}^{\mathrm{p}}/c_{\mathrm{s}} \, \lambda_{\mathrm{mfp}}^{\mathrm{eff}} $ \citep{sharma2003transition, verscharen2017kinetic, kunz2020self}.  Using the wavenumber model (Eq. \eqref{eq:k_anisotropy}), the transition wavenumber is,
\begin{align} \label{eq:k_trans}
k^{\mathrm{trans}} = \frac{v_{\mathrm{th}}^{\mathrm{p}}}{c_{\mathrm{s}} \, \lambda_{\mathrm{mfp}}^{\mathrm{eff}} } \,  \sqrt{1 + \Bigg [\frac{2 \big( v_{\mathrm{th}}^{\mathrm{p}} \big)^2 }{ \big(\lambda_{\mathrm{mfp}}^{\mathrm{eff}} \, k_{\mathrm{iso}} \, c_{\mathrm{s}}\big)^2 } \Bigg]^{(1 -  \alpha)/\alpha} }.
\end{align}
Inserting the combined statistics from Table \ref{tab:table1}, using a typical value of $v_{\mathrm{th}}^{\mathrm{p}} / c_{\mathrm{s}} = \sqrt{1/2}$ for the solar wind, and using the TFF assumption, the transition wavenumber in spacecraft-frame frequency at 1 AU is $\langle V_{\mathrm{SW}} \rangle \, k^{\mathrm{trans}} = f^{\mathrm{trans}} =0.19$ Hz, and $\mathrm{CI}_{f^{\mathrm{trans}}} = [0.046, \, 0.33 ]$ Hz. The uncertainties are propagated from $V_{\mathrm{SW}}$ and the four estimates of $k^{\mathrm{trans}}$ from $C_{\parallel}, \, A_{\parallel}, \, C_{\perp}, \, A_{\perp}$.

\emph{Discussion}.--- \label{sec:discussion}
We have measured the relative non-conservation of the CGL invariants and modeled the behavior with the slow-mode branch of the KMHD-BGK equations to measure the effective mean-free-path of the solar wind protons and the scale dependence of the slow-mode wavenumber anisotropy (Table \ref{tab:table1} reports the statistics of these measurements). The primary result of this Letter is the measured effective proton mean-free-path that is $\sim10^3$ times smaller than the Spitzer-H\"{a}rm mean-free-path ($\lambda_{\mathrm{mfp}}^{\mathrm{SH}} = 1.14 \times 10^8 \, \mathrm{km}$, measured with this dataset). Therefore, the fluidlike range in the solar wind extends to much smaller scales than predicted based on particle collisions. In addition, the scale dependent anisotropy of the compressive fluctuations ($\alpha \simeq 0.4$) is consistent with previous measurements \citep{chen2012three, chen2016recent}, being more anisotropic than the Alfv\'{e}nic fluctuations.

The measured transition frequency, the scale between fluid behavior ($ f \ll f^{\mathrm{trans}}$) and collisionless behavior ($ f \gg f^{\mathrm{trans}}$), of $f^{\mathrm{trans}} = 0.189$ Hz is at the well-known break in power law $(k_{\perp} \rho_{\mathrm{p}} \sim 1)$ of the magnetic field power spectrum at 1 AU \citep{leamon1998observational, kiyani2015dissipation, verscharen2019multi}. These measurements therefore justify the use of fluid MHD theory at larger scales $(k_{\perp} \rho_{\mathrm{p}} < 1)$ \citep{zank1996evolution, matthaeus1982measurement, chen2016recent, montgomery1987density, tu1995mhd, coleman1968turbulence, bruno2013solar, goldreich1997magnetohydrodynamic, matthaeus1999turbulence,stawarz2009turbulent, coburn2012turbulent, macbride2008turbulent, marsch1990radial}. If the result $k_{\perp} \rho_{\mathrm{p}} \simeq k_{\parallel} \lambda^{\mathrm{eff}}_{\mathrm{mfp}}$ turns out to be a general property of weakly collisional plasma, this provides a simple parameterization for the effective collisionality of astrophysical plasmas.

Effective collisional processes have long been studied theoretically and numerically \citep{schekochihin2006turbulence, rincon2016turbulent, helander2016contraints, kunz2016magnetorotational, squire2017pressure, kunz2014firehose, coroniti1977magnetic}, but it is an open question as to the relevant role of the various mechanisms \citep{gary2000electromagnetic, hamasaki1973relaxation, marsch2006kinetic, yoon2017kinetic, meyrand2019fluidization, schekochihin2016phase, kellogg2000fluctuations}) and how they are activated \citep{squire2017kinetic, kunz2020self, verscharen2016collisionless}. Therefore, further studies are necessary to assess exactly what key physics of weakly collisional plasma leads to the measured effective collisionality, since most astrophysical plasmas, being multi-scale and turbulent, will support effective collision mechanisms \citep{zhuravleva2019suppressed}. The measurements presented here provide constraints to be satisfied by theories of effective collision processes.

\nocite{*}

\bibliographystyle{apsrev4-2}
\bibliography{apssamp}

\pagebreak
\pagebreak
\widetext
\begin{center}
\textbf{\large Supplemental Materials: Title for main text}
\end{center}
\setcounter{equation}{0}
\setcounter{figure}{0}
\setcounter{table}{0}
\setcounter{page}{1}
\makeatletter
\renewcommand{\theequation}{S\arabic{equation}}
\renewcommand{\thefigure}{S\arabic{figure}}

\section{Outline of supplemental materials}

This document provides supplemental materials for the Letter titled, ``Measurement of the Effective Mean-Free-Path of the Solar Wind Protons''. Section \ref{sec:Wavenumber model} details the wavenumber model (Eq. 3 of the Letter) and the transition wavenumber (Eq. 6 of the Letter), Section \ref{sec:Weighted Geometric Statistics} introduces the weighted geometric statistics (reported in Table 1 of the Letter) and shows a figure of the weighted geometric statistics, and Section \ref{sec:Collision length and time-scales} introduces the Spitzer-H\"{a}rm collision quantities (referenced in the text of the Letter). Section \ref{sec:Linear Collisional - Kinetic Magnetohydrodynamic} introduces the collisional-kinetic magnetohydrodynamic equations, shows the linear Fourier analysis, normalisation and the linear system of equations to be solved numerically.

\section{Wavenumber model} \label{sec:Wavenumber model}
The model $k_{\parallel} \sim k_{\perp}^{\alpha}$ introduced here is generalized from the critical balance model of Alfv\'{e}nic turbulence (see Ref. \citep{goldreich1995toward}). In the Letter it is used to model the compressive wave propagation angle $\theta_{\hat{{\bf b}}, \hat{{\bf k}}}$. To ensure the isotropic scale is defined correctly we begin with,
\begin{align}
\frac{k_{\parallel}}{k_{\mathrm{iso}} / \sqrt{2} } = \Bigg( \frac{k_{\perp}}{k_{\mathrm{iso}} / \sqrt{2}}  \Bigg)^{\alpha}
\end{align}
where $\alpha$ is the anisotropy exponent. Since, $k_{\parallel} = k \, \mathrm{\mathrm{cos}}(\theta_{\hat{{\bf b}}, \hat{{\bf k}}}), \; k_{\perp} = k \, \mathrm{\mathrm{sin}}(\theta_{\hat{{\bf b}}, \hat{{\bf k}}})$ we have,
\begin{align}
\frac{k \, \mathrm{cos}(\theta_{\hat{{\bf b}}, \hat{{\bf k}}})}{ k_{\mathrm{iso}} /\sqrt{2}} = \Bigg( \frac{k \, \mathrm{sin} (\theta_{\hat{{\bf b}}, \hat{{\bf k}}})}{k_{\mathrm{iso}}/\sqrt{2} }  \Bigg)^{\alpha} 
\; \Rightarrow \; 
k^{1-\alpha} = \bigg( \frac{k_{\mathrm{iso}}}{ \sqrt{2}} \bigg)^{1 - \alpha} \mathrm{\mathrm{cos}}(\theta_{\hat{{\bf b}}, \hat{{\bf k}}})^{-1} \, \mathrm{sin}(\theta_{\hat{{\bf b}}, \hat{{\bf k}}})^{\alpha},
\end{align}
so that at $\theta_{\hat{{\bf b}}, \hat{{\bf k}}}^*  = 45^{\circ}$, $\mathrm{sin}(\theta_{\hat{{\bf b}}, \hat{{\bf k}}}^*) = \mathrm{cos}(\theta_{\hat{{\bf b}}, \hat{{\bf k}}}^*) = 1/\sqrt{2}$, we have,
\begin{align}
k^{1-\alpha} =  \bigg( \frac{k_{\mathrm{iso}}}{ \sqrt{2}} \bigg)^{1 - \alpha} \big(\sqrt{2}\big)^{1 - \alpha} = ( k_{\mathrm{iso}})^{1 - \alpha}
\end{align}
so we have recovered the isotropic scale $k = k_{\mathrm{iso}}$ where $k_{\perp} = k_{\parallel}$. Writing the wavenumber model
\begin{align} \label{eq:k_model}
k = \frac{k_{\mathrm{iso}}}{ \sqrt{2}} \, \mathrm{cos}(\theta_{\hat{{\bf b}}, \hat{{\bf k}}})^{1 / (\alpha-1)} \, \mathrm{sin}(\theta_{\hat{{\bf b}}, \hat{{\bf k}}})^{\alpha / (1-\alpha)}.
\end{align}
This Eq. appears as number 3 of the Letter. Here $k$ depends on $\theta_{\hat{{\bf b}}, \hat{{\bf k}}}$ parametrised by $\alpha  \, \in [0,1), \;  k_{\mathrm{iso}}$. Eq. \ref{eq:k_model} can be inverted on $\theta_{\hat{{\bf b}}, \hat{{\bf k}}} \in [0,90^{\circ})$ for $k$.

The wavenumber model is also used in the derivation of Eq. 6 of the Letter. Just above Eq. 6, in the text of the Letter, the relation,
\begin{align} 
k^{\mathrm{trans}}_{\parallel} \lambda_{\mathrm{mfp}}^{\mathrm{eff}} = v_{\mathrm{th}}^{\mathrm{p}} / c_{\mathrm{s}},
\end{align}
is argued to define $k_{\parallel}^{\mathrm{trans}}$, which can be compared to measurements with the full wavenumber $k^{\mathrm{trans}}$. Using the model (Eq. \ref{eq:k_model}) to write,
\begin{align}
\frac{v_{\mathrm{th}}^{\mathrm{p}} }{ c_{\mathrm{s}} \, \lambda_{\mathrm{mfp}}^{\mathrm{eff}} } = k^{\mathrm{trans}} \mathrm{\mathrm{cos}} (\theta_{\hat{{\bf b}}, \hat{{\bf k}}}^{\mathrm{trans}}) = \frac{k_{\mathrm{iso}}}{ \sqrt{2}} \, \mathrm{tan}(\theta_{\hat{{\bf b}}, \hat{{\bf k}}}^{\mathrm{trans}})^{\alpha / (1-\alpha)}.
\end{align}
Solving for $\theta_{\hat{{\bf b}}, \hat{{\bf k}}}^{\mathrm{trans}}$,
\begin{align}
\theta_{\hat{{\bf b}}, \hat{{\bf k}}}^{\mathrm{trans}} = \mathrm{arctan}  \bigg\{ \bigg [\frac{ \sqrt{2} v_{\mathrm{th}}^{\mathrm{p}} }{ c_{\mathrm{s}} \, k_{\mathrm{iso}} \, \lambda_{\mathrm{mfp}}^{\mathrm{eff}} } \bigg]^{(1-\alpha)/\alpha} \bigg \}.
\end{align}
Now, $k^{\mathrm{trans}}$ can be written,
\begin{align}
k^{\mathrm{trans}} = \frac{k_{\mathrm{iso}}}{ \sqrt{2}} \, \mathrm{cos}(\theta_{\hat{{\bf b}}, \hat{{\bf k}}}^{\mathrm{trans}})^{1 / \alpha-1} \, \mathrm{sin}(\theta_{\hat{{\bf b}}, \hat{{\bf k}}}^{\mathrm{trans}})^{\alpha / 1-\alpha}.
\end{align}
Using the trigonometric identities,
\begin{align}
\mathrm{\mathrm{cos}}(\mathrm{arctan}(x)) = \frac{1}{\sqrt{1 + x^2}} , \; \mathrm{\mathrm{sin}}(\mathrm{arctan}(x)) = \frac{x}{\sqrt{1 + x^2}} ,
\end{align}
yields,
\begin{align}
k^{\mathrm{trans}} = \frac{k_{\mathrm{iso}}}{ \sqrt{2}} \, \bigg(\frac{1}{\sqrt{1 + \chi^2}}  \bigg)^{1 / \alpha-1} \, \bigg( \frac{\chi}{\sqrt{1 + \chi^2}}\bigg)^{\alpha / 1-\alpha}.
\end{align}
Where,
\begin{align}
\chi = \Bigg [\frac{ \sqrt{2} v_{\mathrm{th}}^{\mathrm{p}} }{ c_{\mathrm{s}} \, k_{\mathrm{iso}} \, \lambda_{\mathrm{mfp}}^{\mathrm{eff}} } \Bigg]^{(1 - \alpha)/\alpha}
\end{align}
which simplifies to,
\begin{align}
k^{\mathrm{trans}} = \frac{k_{\mathrm{iso}}}{ \sqrt{2}} \, \chi^{\alpha / (1-\alpha)}\sqrt{1 + \chi^2}.
\end{align}
Now inserting $\chi$,
\begin{align}
k^{\mathrm{trans}} = \frac{ v_{\mathrm{th}}^{\mathrm{p}} }{ c_{\mathrm{s}} \, \lambda_{\mathrm{mfp}}^{\mathrm{eff}} } \,  \sqrt{1 + \bigg [\frac{ 2 ( v_{\mathrm{th}}^{\mathrm{p}})^2 }{ c_{\mathrm{s}}^2 \, k_{\mathrm{iso}}^2 \, (\lambda_{\mathrm{mfp}}^{\mathrm{eff}})^2 } \bigg]^{(1 -  \alpha)/\alpha} }.
\end{align}
This appears as Eq. 6 of the letter.

\section{Weighted Geometric Statistics} \label{sec:Weighted Geometric Statistics}
Following Refs. \citep{kendall1946advanced, norris1940standard}. For the $i$th observation $x_i$, the geometric mean is,
\begin{align}
\mu_{\mathrm{g}}^x =  \mathrm{exp} \bigg\{ \frac{1}{n} \sum_i^n \mathrm{ln} | x_i | \bigg\},
\end{align}
where $n$ is the number of observations. It is noted that,
\begin{align}
\mathrm{ln} |\mu_{\mathrm{g}}^x| = \frac{1}{n} \sum_i^n \mathrm{ln} | x_i | = \mathrm{E}_{\mathrm{a}} \big( \mathrm{ln} | x_i | \big),
\end{align}
where the operator,
\begin{align}
\mathrm{E}_{\mathrm{a}} = \frac{1}{n} \sum_i^n,
\end{align}
is the arithmetic expectation value, so that the natural logarithm of the geometric mean is the arithmetic expectation of the natural logarithm of the observations. If the observations have unormalized weights $w_i$, the weighted geometric mean,
\begin{align}
\mu_{\mathrm{w.g.}}^x  =  \mathrm{exp} \bigg\{ \sum_i^n \frac{ w_i \, \mathrm{ln} | x_i | }{ \sum_i^n w_i} \bigg\},
\end{align}
which reduces to the geometric mean if $w_i = 1 \; \forall \;i$. The same applies for the expectation values,
\begin{align} \label{eq:wg_mean}
\mathrm{ln}| \mu_{\mathrm{w.g.}} | =  \sum_i^n \frac{ w_i \, \mathrm{ln} | x_i | }{\sum_i^n w_i} = \mathrm{E}_{\mathrm{w.a.}} \big(\mathrm{ln} | x_i | \big),
\end{align}
where,
\begin{align}
\mathrm{E}_{\mathrm{w.a.}} = \frac{1}{n}  \sum_i^n \, \frac{w_i}{ \sum_i^n w_i},
\end{align}
is the weighted arithmetic expectation operator. The arithmetic covariance matrix is,
\begin{align}
 \big(\sigma_{\mathrm{a}}^{x,y} \big)^2 = \mathrm{E}_{\mathrm{a}} \big[ \big( x_i - \mathrm{E}_{\mathrm{a}}(x_i) \big) \big(y_i - \mathrm{E}_{\mathrm{a}}(y_i)\big)\big],
\end{align}
It follows then that the weighted geometric covariance matrix,
\begin{align}
\mathrm{ln} \big|  \big(\sigma_{\mathrm{w.g.}}^{x,y} \big)^2  \big| = \mathrm{E}_{\mathrm{w.a.}} \big[ \big( X_i - \mathrm{E}_{\mathrm{w.a.}} (X_i) \big) \big( Y_i - \mathrm{E}_{\mathrm{w.a.}} (Y_i) \big) \big],
\end{align}
where $X_i = \mathrm{ln} | x_i |$, giving,
\begin{align} \label{eq:wg_cov_mat}
 \big(\sigma_{\mathrm{w.g.}}^{x,y} \big)^2 = \mathrm{exp} \bigg\{ \frac{1}{\sum_i^n w_i } \sum_i^n w_i \,  \mathrm{ln} \bigg| \frac{x_i}{\mu_{\mathrm{w.g.}}^x}  \bigg| \,   \mathrm{ln} \bigg| \frac{y_i}{\mu_{\mathrm{w.g.}}^y}  \bigg| \bigg\}.
\end{align}
The confidence interval for the arithmetic statistics,
\begin{align} 
\mu_{\mathrm{a}}^x \pm \sigma^{x,x}_{\mathrm{a}} \; \Rightarrow \; \mathrm{CI}_{\mathrm{a}}^{x} = [\mu^x_{\mathrm{a}} - \sigma^{x,x}_{\mathrm{a}} , \,  \mu^x_{\mathrm{a}} + \sigma^{x,x}_{\mathrm{a}}],
\end{align}
where $\mathrm{CI}_{\mathrm{a}}^{x}$ is the arithmetic confidence interval. For the two sigma geometric confidence,
\begin{align} \label{eq:wg_ci}
\mathrm{ln} |\mu^x_{\mathrm{w.g.}} | \pm \mathrm{ln} |(\sigma^{x,x}_{\mathrm{w.g.}})^2| =     
\begin{cases}
      \mathrm{ln} |\mu^x_{\mathrm{w.g.}} \, (\sigma^{x,x}_{\mathrm{w.g.}})^2 |, & + \\
      \mathrm{ln} |\mu^x_{\mathrm{w.g.}} / (\sigma^{x,x}_{\mathrm{w.g.}})^2 |, & -
    \end{cases}
\; \Rightarrow \; \mathrm{CI}_{\mathrm{w.g.}}^{x} = [ \mu^x_{\mathrm{w.g.}} \, (\sigma^{x,x}_{\mathrm{w.g.}})^2 , \, \mu^x_{\mathrm{w.g.}} /( \sigma^{x,x}_{\mathrm{w.g.}})^2 ],
\end{align}
where $\mathrm{CI}_{\mathrm{w.g.}}^{x}$ is the weighted geometric confidence interval. The statistics detailed here are used to calculate the main results of the Letter, which are reported in Table 1. In the Letter, the subscripts w.g. has been dropped.

\begin{figure}
\includegraphics[width=160mm]{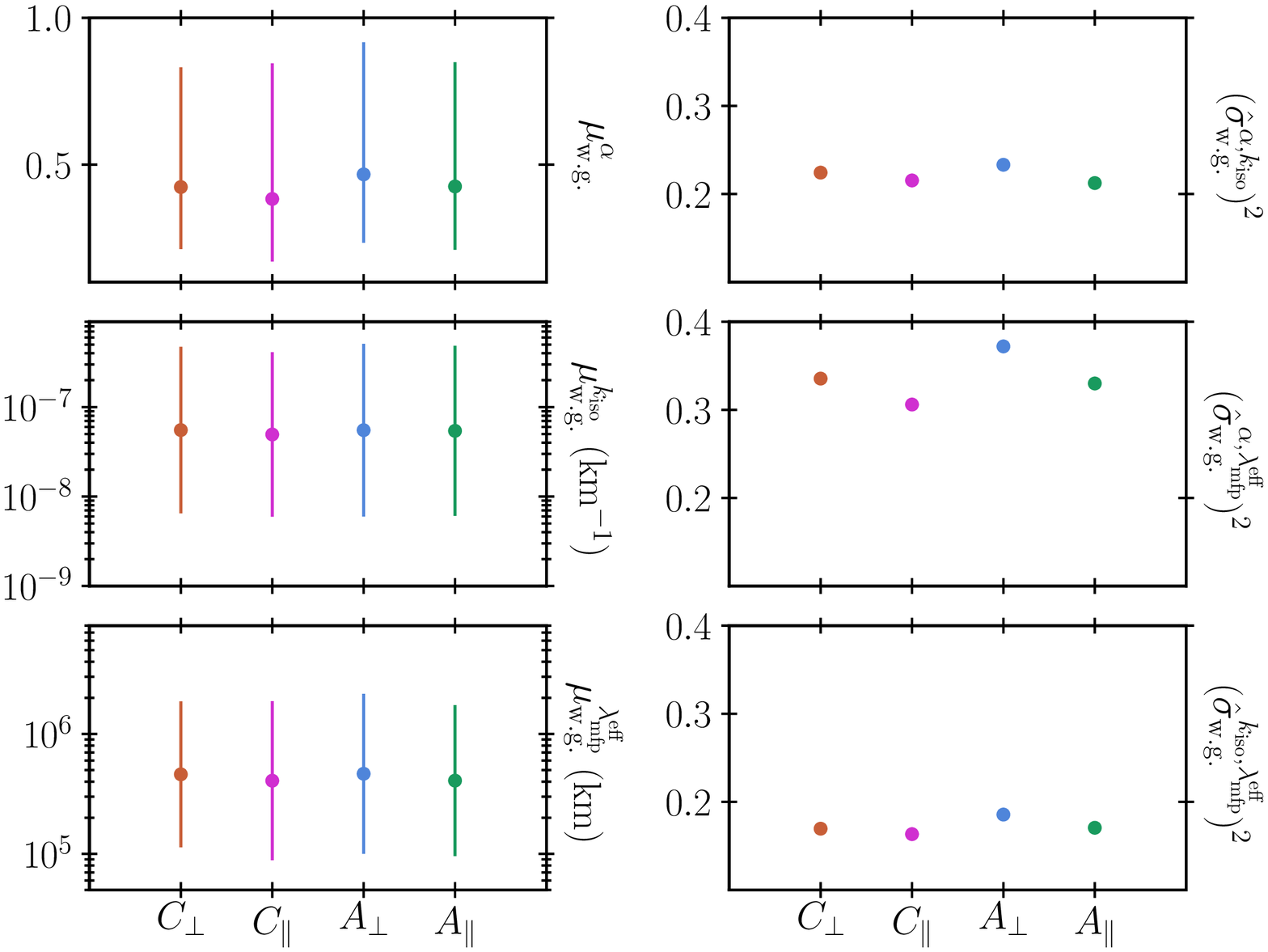}
\caption{\label{fig:geom_stats} The weighted geometric mean (Eq. \ref{eq:wg_mean}) are plotted on the left column as circles. The confidence interval (Eq. \ref{eq:wg_ci}) is plotted as a vertical line. The right column are the non-diagonal terms of the normalised weighted covariance matrix (Eq. \ref{eq:norm_wg_cov_mat}). }
\end{figure} 

In the Letter it is stated that the methods used to measure the three unobserved parameters $\alpha, k_{\mathrm{iso}}, \lambda_{\mathrm{mfp}}^{\mathrm{eff}}$ has been repeated for the measurables $C_{\perp}, C_{\parallel}, A_{\perp}, A_{\parallel}$. The left column of Fig \ref{fig:geom_stats} displays the statistics. The means and confidence intervals are consistent among all the measurements, and so the combined statistics are reported in the Letter. 

The right column of Fig. \ref{fig:geom_stats} shows the normalized weighted geometric covariance between the model parameters,
\begin{align} \label{eq:norm_wg_cov_mat}
\big( \hat{\sigma}^{x,y}_{\mathrm{w.g.}} \big)^2 = \frac{\big( \sigma^{x,y}_{\mathrm{w.g.}} \big)^2}{ \sigma^{x,x}_{\mathrm{w.g.}}  \,  \sigma^{y,y}_{\mathrm{w.g.}} }.
\end{align}
They are consistent with each other, so that only the combined statistics are reported in the Letter.

\section{Collision length and time-scales} \label{sec:Collision length and time-scales}
Following Refs. \citep{book1987nrl, spitzer2006physics} the Spitzer-H\"{a}rm proton-proton collision frequency for a proton-electron plasma with $T_{\mathrm{p}} \leq T_{\mathrm{e}}$, where $T_{\mathrm{p}} \, (T_{\mathrm{e}})$ is the proton (electron) temperature, where the inequality sign is the typical case for the solar wind plasma, is written,
\begin{align}
\nu_{\mathrm{p},\mathrm{p}}^{\mathrm{SH}} = 4.8 \times 10^{-8} \, n_{\mathrm{p}} (k_B T_{\mathrm{p}})^{-3/2} \, \Lambda \; \mathrm{s}^{-1},
\end{align}
where $n_{\mathrm{p}} \, (\mathrm{cm}^{-3})$ is the proton number density, $k_B T_{\mathrm{p}}$ is in eV and the Coulomb logarithm is $ \mathrm{ln} |\Lambda|$. The Coulomb logarithm for proton-proton collisions,
\begin{align}
\Lambda = 23 - \mathrm{ln} \bigg| \frac{\sqrt{2 n_{\mathrm{p}}}}{T_{\mathrm{p}}^{3/2}} \bigg|.
\end{align}
The dataset described in section \emph{Measurements} of the Letter provides the following averages,
\begin{align}
n_{\mathrm{p}} &= 5.33 \, (\mathrm{cm}^{-3}), \\
T_{\mathrm{p}} &= 30.0 \, (\mathrm{eV}), \\
v_{\mathrm{th}}^{\mathrm{p}}  &= 48.3 \, (\mathrm{km/s}),
\end{align}
where the proton thermal speed is $v_{\mathrm{th}}^{\mathrm{p}}$. With these measurements we calculate,
\begin{align}
\nu_{\mathrm{p},\mathrm{p}}^{\mathrm{SH}} &= 4.23 \times 10^{-7} \, (s^{-1}), \\
\lambda_{\mathrm{mfp}}^{\mathrm{SH}} &= v_{\mathrm{th}}^{\mathrm{p}} / \nu_{\mathrm{p},\mathrm{p}}^{\mathrm{SH}} = 1.14 \times 10^{8} \, (\mathrm{km}),
\end{align}
where $\lambda_{\mathrm{mfp}}^{\mathrm{SH}}$ is the Spitzer-H\"{a}rm proton-proton mean-free-path.

\section{Linear Collisional-Kinetic Magnetohydrodynamic} \label{sec:Linear Collisional - Kinetic Magnetohydrodynamic}
In this section the eigenvalue problem that is solved to report all numerical quantities is detailed. The kinetic magnetohydrodynamic equations are found in Ref. \citep{kulsrud1983mhd}. The collision operator is the Bhatnagar-Gross-Krook [BGK] introduced in the research article Ref. \citep{gross1956model}. This form of equations, now called KMHD-BGK, which appears in Ref.  \citep{snyder1997landau}, has been studied previously because they offer a linear closure scheme that incorporates Landau/Barnes damping and collisions or a relaxation processes.  In Ref. \citep{sharma2003transition} they show the dispersion relations produced from the KMHD-BGK equations transition from collisional (fluid) to collisionless wave mode solutions, which is the primary purpose of our study.

The equations are introduced in Section \ref{sec:Overview of equation}, linearized and Fourier analyzed in Section \ref{sec:Linearization and Fourier Analysis}, and written as a linear homogenous system of equations in Section \ref{sec:normalizations}. This can be solved with a numerical treatment of the plasma dispersion function and a numerical algorithm for eigenvalue problems. Solutions to these equations form the numerical treatment in the Letter.

\subsection{Overview of equations} \label{sec:Overview of equation}
Beginning with the Vlasov equations and transforming the coordinates into a new velocity frame $v_i \to w_i = v_i - u_i^s(t, {\bf r})$ shifted by the species ``$s$" bulk velocity $u_i^s(t, {\bf r})$ it can be shown that a gyrotropic distribution function $f_s = f_s({\bf x}, {\bf w}, t)$ will evolve according to,
 \begin{align} \label{eq:Drift kinetic equation text}
\Bigg\{ \frac{\partial }{\partial t} & + \big(\hat{b}_i w_{\parallel} +  u_i^s\big) \frac{\partial }{\partial x_i} + \frac{w^2_{\perp} }{2} \frac{\partial \hat{b}_i}{\partial x_i}   \frac{\partial }{\partial w_{\parallel}}
 \nonumber \\
&   + \bigg[ \frac{q_s}{m_s}  E_{\parallel}    - \hat{b}_i \bigg( \frac{\partial u_i^s}{\partial t} + u_j^s   \frac{\partial u_i^s}{\partial x_j} \bigg) \bigg]   \bigg(\frac{\partial }{\partial w_{\parallel}}  + \frac{ w_{\parallel}}{w} \frac{\partial }{\partial w} \bigg) 
\nonumber \\
& - \frac{w^2_{\perp}}{2w}   \frac{\partial u_i^s}{\partial x_i} \frac{\partial }{\partial w}  + \hat{b}_i  \hat{b}_j  \frac{\partial u_i^s}{\partial x_j} \bigg[  \bigg( \frac{w^2_{\perp} }{2}    - w^2 _{\parallel} \bigg) \frac{ 1}{w} \frac{\partial }{\partial w} - w_{\parallel} \frac{\partial }{\partial w_{\parallel}}   \bigg] \Bigg\}  f_s  = -\nu_{s} \big [ f_s - F_s].
\end{align}
Index notation is used so that the dot product appears as $a_i b_i$ which implies a sum over $i$. The quantities here are the normalized magnetic field vector $\hat{b}_i = b_i / |b_i|$, the parallel (perpendicular) peculiar velocity in the guiding center frame $w_{\parallel} = w_i \hat{b}_i \; \big(w_{\perp} = \sqrt{w_i^2 - w_{\parallel}^2} \,\big)$ and the parallel electric field $E_{\parallel}$. The equilibrium distribution function,
\begin{align}
F_s = n_s  \bigg( \frac{m_s n_s}{2 \pi p_s} \bigg)^{3/2}  \text{exp} \bigg\{ - \frac{m_s n_s}{2 p_s} (v - u^s_{\parallel})^2 \bigg\},
\end{align}
is assumed to be Maxwellian. The quantites here are the density $n_s$, mass $m_s$, and scalar pressure $\delta_{ij} p_{ij}^s$, where $\delta_{ij}$ is the Kronecker delta and $p_{ij}^s$ is the pressure tensor. The right hand side of Eqn. \ref{eq:Drift kinetic equation text} is the BGK operator where $\nu_s$ is the effective collision frequency. Changing variables from $w_{\parallel}, w \to w_{\parallel}, w_{\perp}^2/2 |b_i|$ reveals the more familiar form of the drift kinetic equation.

These equations are accompanied by the 0th and 1st moments of the drift kinetic equation,
\begin{align}
&\frac{\partial}{\partial t} n_s + \frac{\partial}{\partial x_i} n_s u_i^s = 0 \label{eq:continuity}
\\
& m_s n_s \bigg( \frac{\partial}{\partial t} + u_j^s \frac{\partial}{\partial x_j} \bigg) u_i^s  + \frac{\partial}{\partial x_j} p_{ij}^s - q_s n_s \bigg( E_i + \frac{1}{c} \epsilon_{ijk} u_j^s b_k \bigg) = 0, \label{eq:momentum}
\end{align}
where the quantities are charge $q_s$, electric field vector $E_i$, speed of light $c$, and the Levi-Civita tensor $\epsilon_{ijk}$. The momentum equation is summed over species $s = p,e$ (protons, electrons) so that the electric field can be eliminated with the quasi-neutrality assumption $n_p = n_e = n$. After using the smallness of the mass ratio $m_e/m_p$ (electron mass / proton mass) to eliminate electron inertia terms, Eq. \ref{eq:momentum} becomes,
\begin{align}
m_p n \bigg( \frac{\partial}{\partial t} + u_j^p \frac{\partial}{\partial x_j} \bigg) u_i^p  + \frac{\partial}{\partial x_j} \big( p_{ij}^p + p_{ij}^e \big) -  \frac{1}{c} \epsilon_{ijk} J_j b_k  = 0
\end{align}
where the definition of the current arises, $J_i = q_p n_p u_i^p + q_e n_p u_i^e$. The ideal Ohm's Law,
\begin{align}
E_i = \epsilon_{ijk} u_j^p b_k.
\end{align}
Inserting into Faraday's Law and simplifying,
\begin{align} \label{eq:induction}
\frac{\partial}{\partial t} b_i = b_j \frac{\partial}{\partial x_j} u^p_j - \frac{\partial}{\partial x_j} u^p_j b_i
\end{align}
giving the induction equation. Using the normalizations in Section \ref{sec:normalizations} and taking the speed of light to be much larger than the Alfv\'{e}n velocity we can write the current as,
\begin{align}
J_i = \epsilon_{ijk} \frac{\partial}{\partial x_j} b_k,
\end{align}
giving the total momentum equation,
\begin{align} \label{eq:momentum total}
m_p n \bigg( \frac{\partial}{\partial t} + u_j^p \frac{\partial}{\partial x_j} \bigg) u_i^p  + \frac{\partial}{\partial x_j} \big( p_{ij}^p + p_{ij}^e \big) +  \frac{1}{c} \bigg( b_j \frac{\partial}{\partial x_i} b_j -  b_j \frac{\partial}{\partial x_j} b_i \bigg) = 0.
\end{align}
With the quasi-neutrality condition,
\begin{align}
n_p = n_e,
\end{align}
our equations can be closed by linearizing them and taking density and pressure moments of the drift kinetic equation. 

\subsection{Linearization and Fourier Analysis} \label{sec:Linearization and Fourier Analysis}
The perturbations we use,
\begin{align}
&b_i = b'_i + B \hat{b}_i \\
&E_{\parallel} = E'_{\parallel} \\
& u^s_i  = u^{s \prime}_i \\
&f_s = f_s' + F_s \\
& n_s = n_0^s + n'_s \to n^s + n'_s \\
& p_{\perp} = p_{\perp,0}^s + p_{\perp}^{s \prime} \to p_{\perp}^s + p_{\perp}^{s \prime} \\
& p_{\parallel} = p_{\parallel,0}^s + p_{\parallel}^{s \prime} \to p_{\parallel}^s + p_{\parallel}^{s \prime}.
\end{align}
where the primed variables are the fluctuations about the unprimed variables (background). The background electric field has been ordered out and we are in the guiding center frame so the background velocity field is zero. The Fourier ansatz in space and time,
\begin{align}
f'_s(x_{\perp}, x_{\parallel}, v_{\parallel}, v, t) &= \tilde{f}_s(v_{\parallel}, v, t) \, \text{exp}  \{ i ( k_{\perp} x_{\perp} + k_{\parallel} x_{\parallel} - \omega t) \} \\
b'_i(x_{\perp}, x_{\parallel},  t) &= \tilde{b}_i \, \text{exp}  \{ i ( k_{\perp} x_{\perp} + k_{\parallel} x_{\parallel} - \omega t) \} \\
u'_i(x_{\perp}, x_{\parallel},  t) &= \tilde{u}_i \, \text{exp}  \{ i ( k_{\perp} x_{\perp} + k_{\parallel} x_{\parallel} - \omega t) \} \\
E'_{\parallel}(x_{\perp}, x_{\parallel},  t) &= \tilde{E}_{\parallel} \, \text{exp}  \{ i ( k_{\perp} x_{\perp} + k_{\parallel} x_{\parallel} - \omega t) \} \\
\tilde{n}_s(x_{\perp}, x_{\parallel},  t) &= \int d^3w \; \tilde{f}_s(x_{\perp}, x_{\parallel}, w_{\parallel}, w, t) \\
\tilde{p}^s_{\perp}(x_{\perp}, x_{\parallel},  t) &= \frac{m_s}{2} \int d^3v \; w_{\perp}^2 \, \tilde{f}_s(x_{\perp}, x_{\parallel}, w_{\parallel}, w, t) \\
\tilde{p}^s_{\parallel}(x_{\perp}, x_{\parallel},  t) &= m_s \int d^3w \; w_{\parallel}^2 \, \tilde{f}_s(x_{\perp}, x_{\parallel}, w_{\parallel}, w, t)
\end{align}
where the wavenumber $k_i = 
\delta_{i \perp} k_{\perp} + \delta_{i \parallel} k_{\parallel}$ and complex frequency $\omega$ have been introduced. The Fourier amplitude of the perturbed distribution function,
\begin{align} \label{eq:Fourier amplitude distribution function}
& \tilde{f}_s =  \frac{1}{i \big( -  \omega  +  k_{\parallel} w_{\parallel}  - i \nu_{s} \big)} \bigg( \frac{n_s q_s}{p_s} \, \tilde{E}_{\parallel}  w_{\parallel} F_s  - i\omega   \frac{m_s n_s}{p_s} \frac{\tilde{b}_{\parallel}}{B} F_s \frac{w^2_{\perp}}{2} \bigg)   
\nonumber \\
& \hspace{1cm} - \frac{ i\nu_s}{\big( -  \omega  +  k_{\parallel} w_{\parallel}  - i \nu_{s} \big)} \frac{m_s n_s}{p_s} w_{\parallel} \tilde{u}_{\parallel}^s  F_s  - \frac{m_s n_s}{p_s} w_{\parallel} \tilde{u}_{\parallel}^s  F_s  
\nonumber \\
& \hspace{1cm}  + \frac{\nu_{s} }{i \big( -  \omega  +  k_{\parallel} w_{\parallel}  - i \nu_{s} \big)}   F_s \Bigg\{ \frac{\tilde{n}_s}{n_s} \bigg[ \frac{5}{2} - \frac{m_s n_s}{2 p_s} w^2\bigg] +  \frac{\tilde{p}_s}{p_s} \bigg[\frac{m_s n_s}{2 p_s} w^2 - \frac{3}{2} \bigg] +  \frac{m_s n_s}{p_s} w_{\parallel} \tilde{u}_{\parallel}^{s} \Bigg\}.
\end{align}
This will be integrated for the pressure and density fluctuations. The continuity equation (Eqn. \ref{eq:continuity}) is linearized and Fourier analyzed to produce,
\begin{align}
- i \omega \tilde{n} + i n \big( k_{\perp} \tilde{u}^{p}_{\perp} + k_{\parallel} \tilde{u}^{p}_{\parallel} \big)  = 0.
\end{align}
The continuity equation is necessary for only the protons $s = p$. The total momentum equation is linearized, Fourier analyzed, and then projected onto the directions perpendicular and parallel to the magnetic field.,
\begin{align}
&\omega m_p n \tilde{u}^p_{\parallel} =    k_{\parallel} \big( \tilde{p}^{p}_{\parallel} + \tilde{p}^{e}_{\parallel} \big)
\\
&m_p  n \omega \tilde{u}^p_{\perp} = k_{\perp} \tilde{p}^{p}_{\perp}  + k_{\perp} \tilde{p}^{e}_{\perp}  + \frac{B }{4 \pi} \big(  k_{\perp} \tilde{b}_{\parallel}  -  k_{\parallel}  \tilde{b}_{\perp} \big )
\end{align}
where the result here is for a Maxwellian background distribution function (isotropic). The velocity components are,
\begin{align}
u_{\parallel}^s = \hat{b}_i u_i^s , \; u_{\perp}^s = | u^s_j \big( \delta_{ij} u_i - \hat{b}_i \hat{b}_j \big) |,
\end{align}
and the gyrotropic pressure,
\begin{align}
&p^s_{ij} = p_{\perp}^s (\delta_{ij} - \hat{b}_i \hat{b}_j) + p_{\parallel}^s \hat{b}_i \hat{b}_j
\\
&p^s_{\perp} = \frac{1}{2} p^s_{ij} (\delta_{ij} - \hat{b}_i \hat{b}_j) , \; p_{\parallel}^s = p^s_{ij}  \hat{b}_i \hat{b}_j.
\end{align}
Following the same procedure on the induction equation,
\begin{align}
&\omega \tilde{b}_{\perp} = - B k_{\parallel} \tilde{u}^p_{\perp}
\\
&\omega \tilde{b}_{\parallel} =  B k_{\perp} \tilde{u}^p_{\perp},
\end{align}

To close the system of equations we need equations for $\tilde{p}^s_{\perp}, \, \tilde{p}^s_{\parallel}, \, \tilde{n}_s$ for protons and electrons which will be produced by taking the appropriate moments of Eqn. \ref{eq:Fourier amplitude distribution function}. Taking the moments and rearranging produces,
\begin{align}
&\frac{\tilde{n}_s}{n_s}  - \frac{\tilde{p}_{\perp}^s}{p_s} =    \frac{\omega}{|k_{\parallel}|  v_s}  \frac{\tilde{b}_{\parallel}}{B}  Z(\zeta_s)   - i \frac{\nu_{s}}{v_s |k_{\parallel}| }  \frac{\tilde{n}_s}{n_s} Z(\zeta_s)  + i \frac{\nu_{s}}{v_s |k_{\parallel}| }  \frac{\tilde{p}_s}{p_s}   Z(\zeta_s) 
\\
&[1 + 2  \zeta_s^2  \mathcal{R}(\zeta_s)] \frac{\tilde{n}_s}{n_s} - \mathcal{R}(\zeta_s) \frac{\tilde{p}_{\parallel}}{p_s} =  \frac{\omega}{|k_{\parallel}|  v_s}  \frac{\tilde{b}_{\parallel}}{B}   [ 2 \zeta_s \mathcal{R}(\zeta_s) - Z(\zeta_s)] 
\nonumber \\
& \hspace{1cm}+ i \frac{3}{2} \frac{\nu_{s}}{v_s |k_{\parallel}| }  \frac{\tilde{n}_s}{n_s} \big[ 2 \zeta_s \mathcal{R}(\zeta_s) -  Z(\zeta_s) \big]+ i \frac{1}{2}  \frac{\nu_{s}}{ |k_{\parallel}| v_s}  \bigg( \frac{2}{3} \frac{\tilde{p}_{\perp}^s}{p_s} + \frac{1}{3}  \frac{\tilde{p}_{\parallel}^s}{p_s} \bigg) \big[ Z(\zeta_s) - 2 \zeta_s \mathcal{R}(\zeta_s)  \big] 
\end{align}
where the plasma dispersion function is written as $Z(\zeta_s)$ appearing in the quantity $\mathcal{R}(\zeta_s) = 1 + \zeta_s Z(\zeta_s)$. The argument,
\begin{align}
\zeta_s = \frac{1}{k_{\parallel} v_s} \bigg( \omega + i \nu_s \bigg),
\end{align}
where $v_s$ is the species thermal speed, $\zeta_s$ is the location of the pole in complex space.

\subsection{Normalization and eigen-equations} \label{sec:normalizations}
Now we introduce the normalizations
\begin{align}
& \nu \to \nu \, \omega_p, \; \tilde{n} \to \tilde{n} \, n_0, \; \tilde{u}_{\alpha} \to \tilde{u}_{\alpha} v_s, \; \omega \to \omega \, \omega_p,
 \nonumber \\
 & \tilde{b}_{\alpha} \to \tilde{b}_{\alpha} \, B, \; \tilde{p}^s_{\alpha} \to \tilde{p}^s_{\alpha} \, p_B, \; k_{\alpha} \to k_{\alpha} v_{s}^{-1} \omega_p 
\end{align}
where $\alpha = \perp, \parallel$ and we introduced the plasma frequency $\omega_p = (4 \pi n q_s^2 / m_p)^{1/2}$, the Alfv\'{e}n speed $v_a = B / (4 \pi n m_p)^{1/2}$, the thermal speed $v_s = ( 2 k_B T_s / m_s)^{1/2}$ and the magnetic pressure $p_B = B^2 / 8 \pi$. The temperature is defined $T_s = p_s / n_s k_B$. The proton plasma beta $\beta^p = v_p^2 / v_A^2 = p^p / p_B = 8 \pi n k_B T_p / B^2$ where, $p^s = (2 p_{\perp}^s + p_{\parallel}^s)/3$ is the background total pressure. Rewriting the equations we obtain,
\begin{align}
& \omega \tilde{u}_{\perp} + \frac{1}{\beta_p} \big( - k_{\perp} \tilde{b}_{\parallel} +  k_{\parallel}  \tilde{b}_{\perp} \big)  - \frac{ k_{\perp}}{2 \beta_p} \big( \tilde{p}_{\perp}^p + \tilde{p}_{\perp}^e)  = 0
\\
& \omega \tilde{u}_{\parallel} -  \frac{k_{\parallel} }{2 \beta_p} \tilde{p}_{\parallel}  = 0
\\
& \omega \tilde{b}_{\perp} +   k_{\parallel}  \tilde{u}_{\perp} = 0
\\
& \omega \tilde{b}_{\parallel} -  k_{\perp}\tilde{u}_{\perp} = 0
\\
& \omega \tilde{n} -  \big( k_{\parallel} \tilde{u}_{\parallel} + k_{\perp} \tilde{u}_{\perp} \big)  = 0
\end{align}
and the complex poles,
\begin{align}
&\zeta^p  =  \frac{\omega + \nu^p}{ |k_{\parallel}|  v_{th}^p} \to  \frac{\omega + \nu^p}{ |k_{\parallel}| }
\\
&\zeta^e  =  \frac{\omega + \nu^e}{ |k_{\parallel}|  v_{th}^e} \to  \frac{\omega + \nu^e}{ |k_{\parallel}| } \sqrt{\frac{T_0^p m_e}{T_0^e m_p}}
\end{align}
The pressure normalizes as $\tilde{p}^s_{\alpha} / p^s \to \tilde{p}^s_{\alpha} /  \beta^s$ and the rest of the quantities are already normalized so we obtain,
\begin{align}
&\tilde{n} \big( 1 +  \zeta^s_{\nu} \, \mathcal{Z}^s \big) - \frac{\tilde{p}^s_{\perp}}{\beta^s} \bigg(1 + \frac{2}{3}  \zeta^s_{\nu} \, \mathcal{Z} \bigg) - \tilde{b}_{\parallel} \zeta_{\omega} \, \mathcal{Z}^s   - \frac{\tilde{p}^s_{\parallel}}{\beta^s} \zeta^s_{\nu} \, \mathcal{Z}^s \frac{1}{3}  = 0
\\
&\tilde{n} \bigg(1 + 2 (\zeta^s)^2 \mathcal{R}^s  +  \frac{3}{2}\zeta^s_{\nu} \big( \mathcal{Z}^s - 2 \zeta^s \mathcal{R}^s \big) \bigg)   - \frac{\tilde{p}^s_{\parallel}}{\beta^s}  \bigg( \mathcal{R}^s + \frac{1}{6} \zeta^s_{\nu} \big( \mathcal{Z}^s - 2 \zeta^s \mathcal{R}^s \big) \bigg) 
\nonumber \\
& \hspace{3cm}  - \tilde{b}_{\parallel} \bigg( 1 + 2 (\zeta^s)^2 \mathcal{R}^s - \mathcal{R}^s + \zeta^s_{\nu} \big( \mathcal{Z}^s - 2 \zeta \mathcal{R}^s \big) \bigg)  -  \frac{\tilde{p}^s_{\perp}}{\beta^s} \frac{1}{3} \zeta^s_{\nu} \big( \mathcal{Z}^s - 2 \zeta \mathcal{R}^s \big)  = 0
\end{align}
Now we can write the linear system of equations,
\begin{align}
\begin{bmatrix}
\omega & 0 & k_{\parallel}/\beta_p  & -k_{\perp} / \beta_p & 0 & - k_{\perp}/2 \beta_p & 0  & - k_{\perp}/2 \beta_p  & 0 \\
0 &  \omega & 0 & 0 & 0 & 0 & -  k_{\parallel} /2 \beta_p   & 0  & -  k_{\parallel} /2 \beta_p\\
k_{\parallel}  & 0 & \omega &  0 & 0  & 0 & 0   & 0  & 0 \\
- k_{\perp}  & 0 & 0 & \omega & 0 & 0 & 0   & 0  & 0 \\
- k_{\perp} & - k_{\parallel}  & 0 & 0 & \omega & 0 & 0   & 0  & 0 \\
0 &  0 & 0 & \mathcal{A}^p_{64} & \mathcal{A}^p_{65} & \mathcal{A}^p_{66} & \mathcal{A}^p_{67}   & 0  & 0 \\
0 &  0 & 0 & \mathcal{A}^p_{74} & \mathcal{A}^p_{75} & \mathcal{A}^p_{76} & \mathcal{A}^p_{77}   & 0  & 0 \\
0 &  0 & 0 & \mathcal{A}^e_{64} & \mathcal{A}^e_{65}  & 0  & 0 & \mathcal{A}^e_{66} & \mathcal{A}^e_{67}   \\
0 &  0 & 0 & \mathcal{A}^e_{74} & \mathcal{A}^e_{75}  & 0  & 0 & \mathcal{A}^e_{76} & \mathcal{A}^e_{77}   
\end{bmatrix}
\begin{bmatrix}
\tilde{u}_{\perp} \\
\tilde{u}_{\parallel} \\
\tilde{b}_{\perp} \\
\tilde{b}_{\parallel} \\
\tilde{n} \\
\tilde{p}^p_{\perp} \\
\tilde{p}^p_{\parallel} \\
\tilde{p}^e_{\perp} \\
\tilde{p}^e_{\parallel}
\end{bmatrix}
=
\begin{bmatrix}
0 \\
0 \\
0 \\
0 \\
0 \\
0 \\
0 \\
0 \\
0
\end{bmatrix}
\end{align}
where we have defined,
\begin{align}
&\mathcal{A}^s_{64} =  - \zeta_{\omega} \mathcal{Z}^s\\
&\mathcal{A}^s_{65} = 1 + \zeta^s_{\nu} \mathcal{Z}^s \\
&\mathcal{A}^s_{66} =  - \frac{1}{\beta^s} \bigg(1 + \frac{2}{3} \zeta^s_{\nu} \mathcal{Z}^s \bigg) \\
&\mathcal{A}^s_{67} =  - \frac{\zeta^s_{\nu} \mathcal{Z}^s}{3 \beta^s}\\
&\mathcal{A}^s_{74} =  - \bigg( 1 + 2 (\zeta^s)^2 \mathcal{R}^s - \mathcal{R}^s + \zeta^s_{\nu} \big( \mathcal{Z}^s - 2 \zeta^s \mathcal{R}^s \big) \bigg) \\
&\mathcal{A}^s_{75} = \bigg(1 + 2 (\zeta^s)^2 \mathcal{R}^s  +  \frac{3}{2}\zeta^s_{\nu} \big( \mathcal{Z}^s - 2 \zeta^s \mathcal{R}^s \big) \bigg) \\
&\mathcal{A}^s_{76} =   - \frac{1}{3\beta^s} \zeta^s_{\nu} \big( \mathcal{Z}^s - 2 \zeta^s \mathcal{R}^s \big) \\
&\mathcal{A}^s_{77} =  - \frac{1}{\beta^s}  \bigg( \mathcal{R}^s + \frac{1}{6} \zeta^s_{\nu} \big( \mathcal{Z}^s - 2 \zeta^s \mathcal{R}^s \big) \bigg) 
\end{align}
This system of equations can be solved with a numerical recipe for the plasma dispersion function and a numerical root finder. This system of equations provide the numerical solutions used in the analysis that provide the main results of the Letter.

The electron response is modelled with the kinetic equation. The ratio of the species mean-free-paths arises when parametrising this model. In the Letter it is quoted to be $\lambda_{\mathrm{mfp}}^{\mathrm{eff}} / \lambda_{\mathrm{mfp, electrons}}^{\mathrm{eff}} = 5$. This reflects the fact the electron gyroscale is much smaller that the protons. Regarding the numerical code, when increasing the ratio $\lambda_{\mathrm{mfp}}^{\mathrm{eff}} / \lambda_{\mathrm{mfp, electrons}}^{\mathrm{eff}}$ from a value of 1, the dispersion relations do not change drastically, but by about 7, the root finding algorithm begins to fail or the null space of the matrix is not small. Therefore, we kept the parameter to be 5.


\end{document}